\documentstyle[epsfig,longtable]{aipproc}

\begin{document}  

\title{The Full Re-Ionization of Helium}
 
\author{James W. Wadsley, Craig J. Hogan, and Scott F. Anderson}
\address{University of Washington\\
Department of Astronomy\\
Box 351580\\
Seattle, WA 98195-1580}

\maketitle

\begin{abstract}
Observations of resolved HeII Lyman alpha
absorption in spectra of two QSO's suggest that the epoch
of helium ionization occurred at $z\approx 3$.
Proximity zones in the spectra of the quasars ($z=3.18,3.285$)
at 304 {\AA} resemble Stromgren spheres,
 suggesting that the intergalactic medium  is only
 singly ionized in helium.  We present models 
of the proximity effect which include the full physics
of the ionization, heating and cooling
 and an accurately simulated inhomogeneous gas distribution.
In these models the underdense
intergalactic medium 
is heated to at least 10,000-20,000 K after
cooling to as low as a few 1000K due to cosmological expansion,
with higher temperatures achieved farther away from
the quasar due to absorption-hardened ionizing spectra.
The  quasars
turn on for a few $\times 10^7$ years  with a fairly
steady flux output at 228 {\AA}  comparable to the 304 {\AA}  flux output
directly observed with HST.
The recoveries in the spectra occur naturally
due to voids in the IGM and  may provide a fairly model-independent probe
of the baryon density.
\end{abstract}

\begin{figure} 
\centerline{\epsfig{file=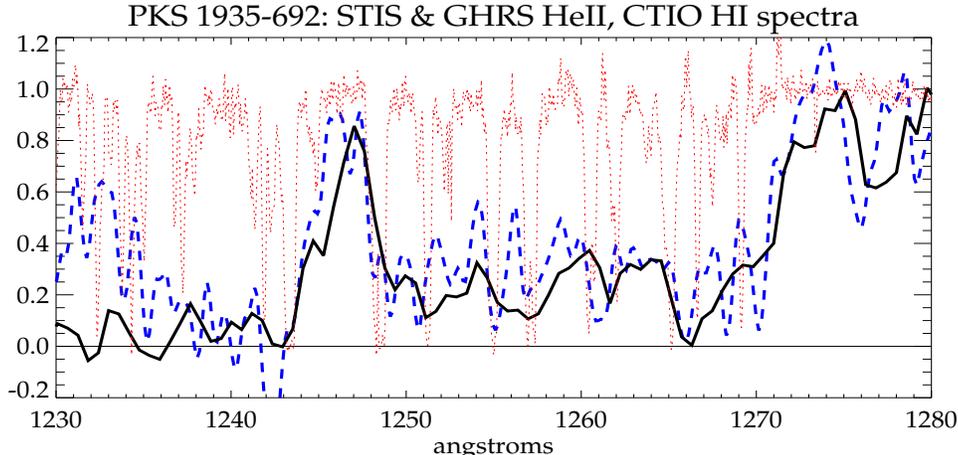,width=5.5in,height=2.5in}}
\vspace{-.1in}
\caption{
Observations of the HeII forest
 of PKS 1935-692 by Anderson et al. (1998) overlaid with a redshift-matched
HI spectrum (dotted).  The shelf structure in the GHRS
(dashed)/STIS (solid) HeII spectra is best explained as the 
second-ionization of helium in the zone near the quasar.
The flux recovers due to a void in the IGM at $\approx 1247${\AA}.
 A similar
pattern is found in the two other higher-$z$ HeII quasars.
}\label{data}
\vspace{-.2in}
\end{figure}

In the last few years it has become possible to observe  details
of absorption by singly ionized helium. The observations combine new
information about the history of quasars, intergalactic gas,
and structure formation. These phenomena  can be disentangled
with detailed quantitative models of the situation which we
briefly describe here.  Theoretical treatments of
 some of the effects modeled here were given by Zheng and Davidsen (1995),
Croft et al. (1997),   Miralda-Escud\'e et al. (1996), Zhang et al. (1998) 
and Fardal, Giroux and Shull (1998).

Early observations of the helium II Lyman alpha absorption spectral
region included the quasars Q0302-003 (z=3.285, Jakobsen et al. 1994),
HS 1700+64 (z=2.72, Davidsen et al. 1996) and PKS 1935-692
(z=3.18, Tytler \& Jakobsen 1996).  Higher resolution (GHRS)
observations of Q0302-003 (Hogan, Anderson \& Rugers 1997) and HE
2347-4342 (z=2.885, Reimers et al. 1997) revealed structure in the
absorption which could be reliably correlated with HI absorption.  The
most recent published observations of PKS 1935-692 with STIS (Anderson
et al. 1998) yield particularly good zero level estimates important
for estimating the optical depth $\tau$. Taken together, these data
now appear to be showing the cosmic ionization of helium by quasars.

All of the objects show absorption with $\tau {{ {\raisebox{-.9ex}{$>$}}\atop{\sim} }} 1$ at
redshifts lower than the quasar.  For the higher redshift QSO's
Q0302-033 and PKS 1935-692
(shown in Figure 1) there is a clear shelf of $\tau {{ {\raisebox{-.9ex}{$>$}}\atop{\sim} }} 1.3$ in a
wavelength region of order 20 {\AA}  in observed wavelength blueward of
the quasar emission line redshift, dropping to a level
 consistent with zero flux or
$\tau {{ {\raisebox{-.9ex}{$>$}}\atop{\sim} }} 3-4$ beyond that.
Anderson et al. conclude from these observations that gas initially 
containing helium as mostly HeII is being  double-ionized in a region
around the quasars.  The lack of a strong emission line for HeII Lyman
alpha suggests  that ionizing flux is escaping
so that the 228 {\AA}   flux may be  similar to a simple
power-law extension of the observed 304 {\AA}   rest frame flux.
Hogan et al. used this to estimate the time required for quasars to
create the double ionized helium region to be 20 Myr for a 20 {\AA}
shelf (dependent on the Hubble parameter, spectral hardness,
cosmology, baryon density and the shelf size). 

The features present in HeII Lyman alpha spectra are reflected in the
HI Lyman alpha forest for these quasars.  Attempts to model the HeII
absorption with line systems detected in HI suggest that low column HI
absorbers, difficult to differentiate from noise in HI spectra,
provide a substantial contribution to the HeII absorption.  Typically,
in the shelf region, the ratio of HeII to HI ions is of order 20 
or more, rising to at least 100 
 farther away (The cross-section for HeII Lyman alpha
absorption is 1/4 that of HI).  A dominant feature in PKS 1935-692 
and HE 2347-4342 is a void in the HeII absorption near the apparent
edge of the HeIII bubble with corresponding voids in the HI
spectra.

To interpret the rich datasets we are constructing models which
include a realistic inhomogeneous distribution of gas as well as the
relevant gas and radiation physics.  We measured density and
temperature along lines of sight through a SPH/N-body cosmological
simulation (Wadsley \& Bond 1998, CDM, $\Omega_b=0.05$, $h=0.5$) to
use for modelling the radiative transfer of 54.4eV radiation from a
newly turned on quasar.  Very small systems produce significant HeII
absorption features, prohibiting using large, poor resolution
simulations.  We generated the long line of sight by bouncing a ray
inside a 5 Mpc comoving diameter typical, mean density simulation.
There is thus no independent long wavelength structure in the spectra.

The quasar flux used was the power-law extension to 228 {\AA}   of the
observed 304 {\AA}   rest frame flux of PKS 1935-692.  There is only
significant continuum
 absorption by HeII when it is the dominant form of helium.
We track  the radiation above 54.4eV (the ionization energy of HeII
to HeIII) in 100 frequency bins.  This is important because the
ionization cross-section for falls off strongly with frequency
as $\nu^{-3}$.  The radiation field thus becomes harder as it is
absorbed moving away from the quasar. 

The gas density was fixed at each point and hydrodynamic motions
ignored, appropriate because of the rapid onset 
of ionization compared to hydrodynamical timescales.
Non-equilibrium energy and ionization equations are evolved
with all the heating, cooling and ionization processes required for a
zero metallicity intergalactic medium: ionization heating,
cosmological expansion, compton, bremsstrahlung, line cooling,
radiative recombination, photoionization and collisional ionization.
Shocks are a possible heating source but the time
scales are sufficiently short that heating associated with 
bulk ionization is dominant.

\begin{figure}[t] 
\centerline{\epsfig{file=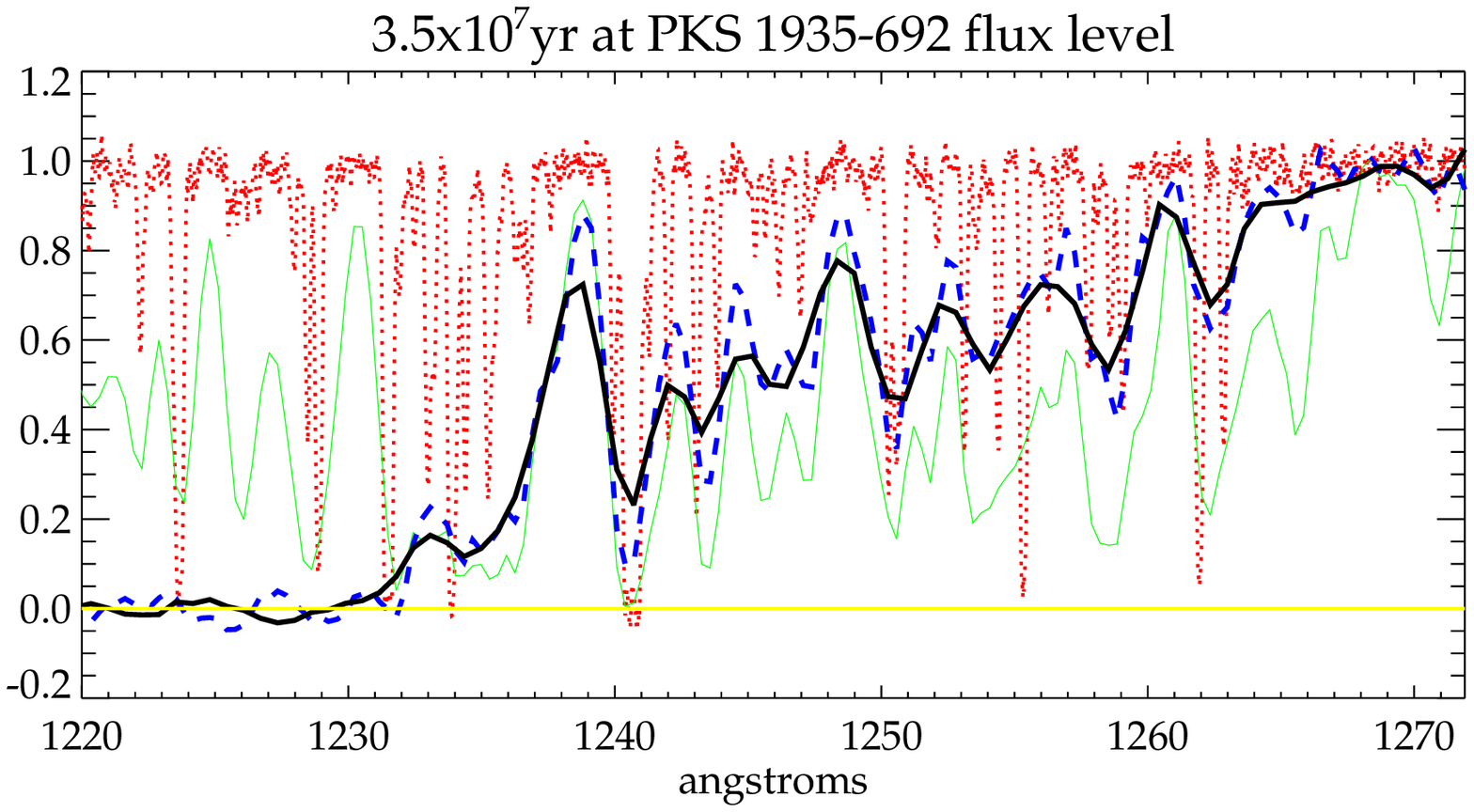,width=5.5in,height=2.5in}}
\vspace{-.2in}
\centerline{\epsfig{file=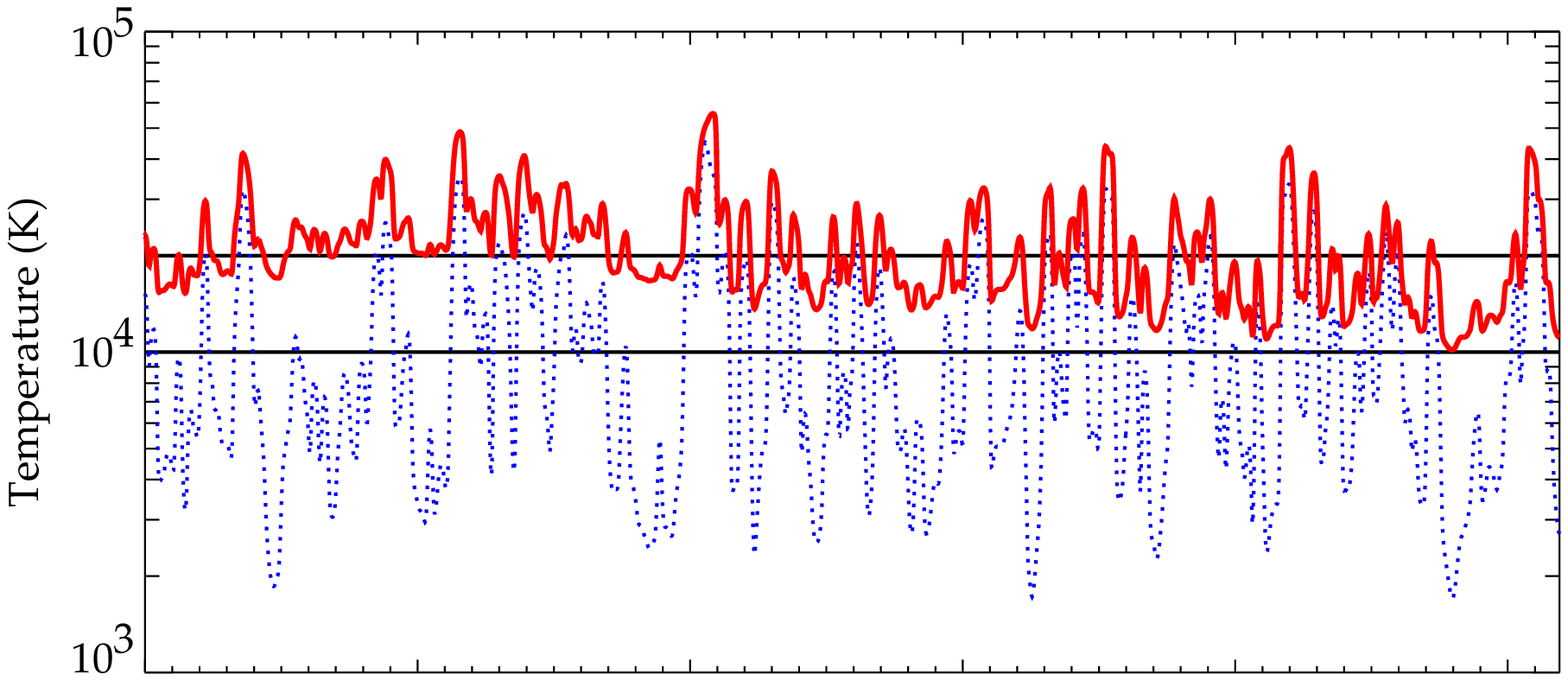,width=5.5in,height=1.6in}}
\vspace{-.2in}
\caption{
Simple model with PKS 1935-692 flux level at 228 {\AA}  and 35 Myr lifetime.  
Simulated spectra: GHRS (dashed), STIS (solid), HI (dotted) and HI
optical depth times 25 (thin solid). The shelf and recovery resemble
those found in the real data, but the Gunn-Peterson edge is
not as pronounced; this can be fixed by allowing for 
quasar variability. The temperature before and after turn-on is 
shown in the lower panel; note the rise in temperature away
from the quasar, especially the order-of-magnitude
increase in the voids.
 The heated zone extends well beyond the edge of
the detectable HeIII bubble.
}\label{model}
\vspace{-.2in}
\end{figure}
 
We treated several lines of sight from the simulation and varied the
flux history and baryon density.  For a given baryon density, the key
parameters are the integrated total luminosity from the quasar,
determining the bubble size, and the flux level for the last few times
$10^6$ years (the recombination time) before the observation is taken,
determining $\tau$.  

A typical model spectrum is shown in Figure 2.  The basic features
of the observed proximity shelves are straightforward to reproduce.
To get substantial recovery ($\tau << 1$) in the spectrum a combination
of a high flux and an empty void is required.  Voids on the edge of
the proximity shelf occur with sufficient frequency in the simulations that
the ubiquity in the observations is not surprising.  The 10,000-20,000K
heating in the medium is greater away from the quasar (lower panel in
Figure 2).  The recombination rate goes as $\sim T^{-0.7}$ and thus
distant voids are made emptier.  This heating and ionization extends
beyond the visible shelf by around $30\%$ beyond which even harder photons get
absorbed.

The time-averaged quasar fluxes could be substantially different from those
observed.  Quasars are known to vary by a factor of two over a
period of  years and the response times are order of $10^5$ years.
Observational bias favours selection of quasars currently at the
bright end of their intrinsic variability.  There is a trade off so
that a higher flux can be used with a corresponding increase in the
baryon density.  Shelves resembling the real data can be 
constructed by suitably adjusting the lifetime and recent  flux.
If quasars are rare density peaks, the mean density nearer quasars is
higher which will increase $\tau$ near the quasar and improve the
model fit.  
  However the density in void regions is seldom less than
$\sim 0.1$ times the cosmological mean and thus we do not have the
freedom to increase the universal baryon fraction without raising the
ionizing flux to compensate, lest the optical depth in the voids become
too high. The voids might therefore offer a relatively model-insensitive
constraint on  mean baryon density.

\begin{figure} 
\centerline{\epsfig{file=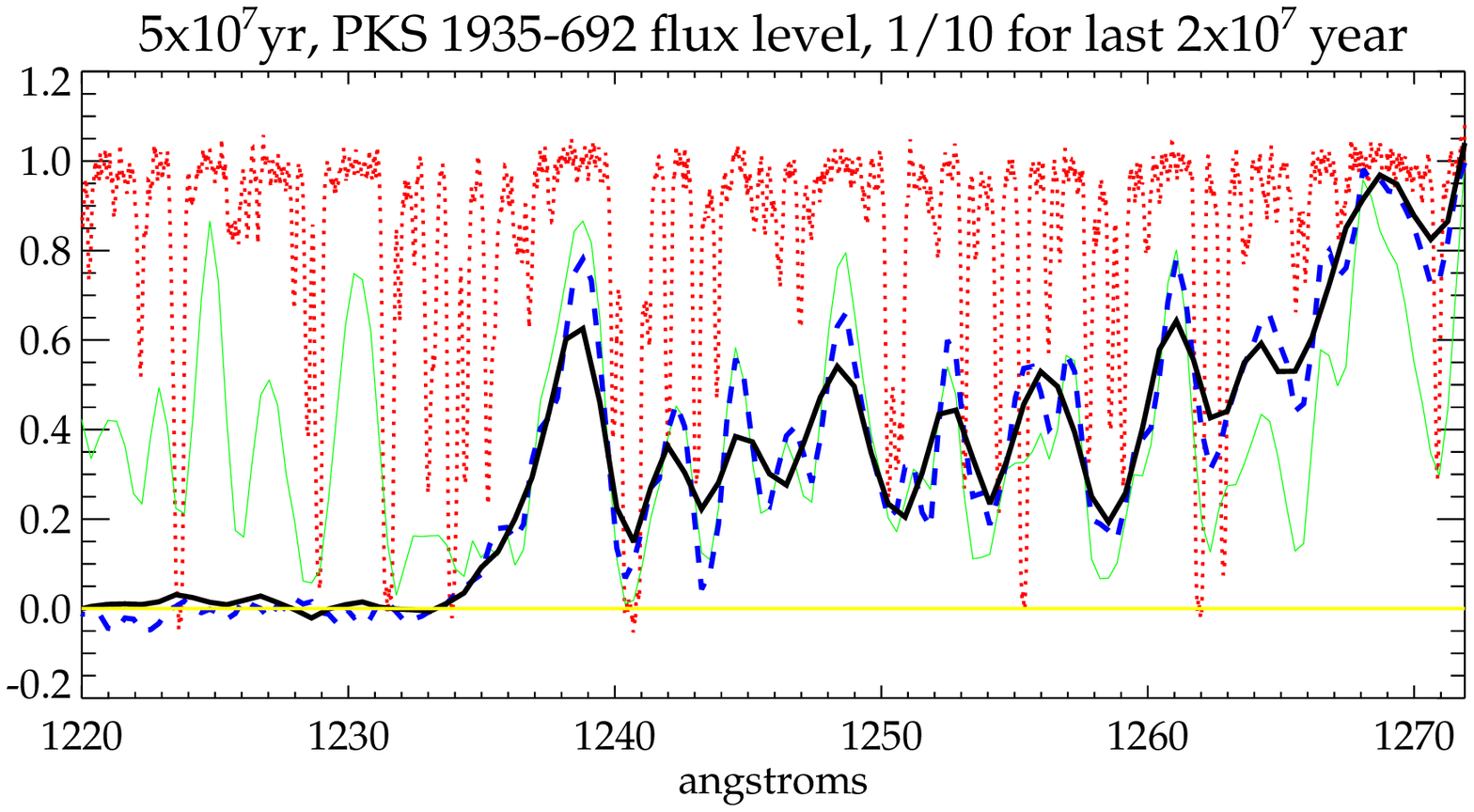,width=5.5in,height=2.5in}}
\vspace{-.2in}
\centerline{\epsfig{file=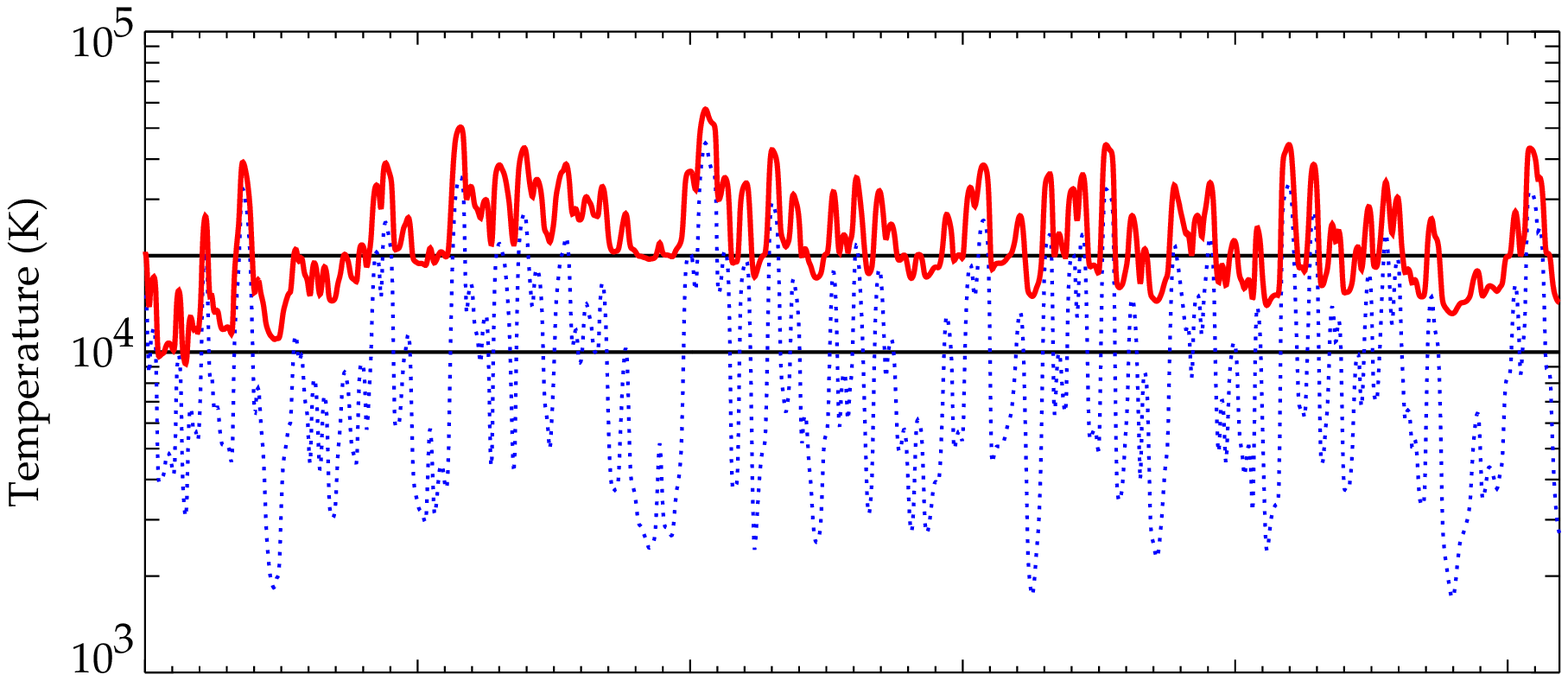,width=5.5in,height=1.6in}}
\vspace{-.2in}
\caption{
Model with PKS 1935-692 flux level at 228 {\AA}  and 50 Myr lifetime with
last 20 My at 1/10 the observed flux.  The curves represent the same
quantities as in Figure 2.
Note that the shelf is better reproduced 
with this choice of parameters.  
The void near 1240 {\AA}  is not as
pronounced.  The intrinsic variability of quasars
allows for some flux variation; but a better way to make the
model match that data is to include the density
enhancement in the zone surrounding the quasar density peak to
increase the shelf optical depth.  The void is sufficiently far from
the quasar to be outside the local density enhancement for reasonable
models of the density peak that produced the quasar host.
This void is at a density 0.11 times the mean which is close to
the rough lower limit of 0.1 times the mean.  This universal
lower limit combined with more detailed modelling currently underway 
will provide a constraint on the mean baryon density.
}\label{model2}
\end{figure}
 
\end{document}